 \documentstyle[12pt]{article}

 \def\bd{\begin{document}} \def\ed{\end{document}}
\def\ds{\documentstyle} \let\fr=\frac \let\bl=\bigl \let\br=\bigr
\let\Br=\Bigr \let\Bl=\Bigl
\let\bm=\bibitem
\let\na=\nabla
\let\pa=\partial \let\ov=\overline
\newcommand{\be}{\begin{equation}}
\newcommand{\ee}{\end{equation}}
\def\ba{\begin{array}}
\def\ea{\end{array}}
\newcommand{\ho}[1]{$\, ^{#1}$}
\newcommand{\hoch}[1]{$\, ^{#1}$}
\newcommand{\bea}{\begin{eqnarray}}
\newcommand{\eea}{\end{eqnarray}}
\newcommand{\ra}{\rightarrow}
\newcommand{\lra}{\longrightarrow}
\newcommand{\Lra}{\Leftrightarrow}
\newcommand{\ap}{\alpha^\prime}
\newcommand{\bp}{\beta^\prime}
\newcommand{\tr}{{\rm tr} }
\newcommand{\Tr}{{\rm Tr} }
\newcommand{\NP}{Nucl. Phys. }
\newcommand{\tamphys}{\it Center for Theoretical Physics\\
Physics Department \\ Texas A \& M University
\\ College Station, Texas 77843}

\newcommand{\auth}{ M. J. Duff{}
\hoch{\ddagger}}

\begin{document}

\hfill{CTP-TAMU-06/93}

\vspace{24pt}

\begin{center}
{ \large {\bf Twenty Years of the Weyl Anomaly\hoch{\dagger}}}

\vspace{36pt}

\auth

\vspace{10pt}

\vspace{10pt}

{\tamphys}

\vspace{44pt}

\underline{ABSTRACT}

\end{center}

In 1973 two Salam prot\'{e}g\'{e}s (Derek Capper and the author) discovered
that the conformal invariance under Weyl rescalings of the metric tensor
$g_{\mu\nu}(x)\rightarrow\Omega^2(x)g_{\mu\nu}(x)$ displayed by classical
massless field systems in interaction with gravity no longer survives in the
quantum theory.  Since then these Weyl anomalies have found a variety of
applications in black hole physics, cosmology, string theory and statistical
mechanics.  We give a nostalgic review.

\vfill \leftline{CTP/TAMU-06/93}\leftline{July 1993}
 \vskip 10pt
\footnoterule
{\footnotesize

\hoch{\dagger}Talk given at the {\it Salamfest}, ICTP, Trieste, March 1993.
\newline
\hoch{\ddagger} Research supported in part by NSF Grant PHY-9106593.
 \vskip -12pt}
\baselineskip=24pt

\pagebreak

\setcounter{page}{1}
\vspace{1cm}
{\it When all else fails, you can always tell the truth.}

{}~~~~~~~~~~~~~~~~~~~~~~~~~~~~~~~~~~~~~~~~~~~~~~~~~~~~~~~~~~~~~~~~~~~~~~Abdus
Salam
\vspace{1cm}
\section{Trieste and Oxford}
\label{Trieste}
 Twenty years ago, Derek Capper and I had embarked on our very first postdocs
here in
Trieste. We were two Salam students fresh from Imperial College filled with
ideas about quantizing
the gravitational field: a subject which at the time was pursued only by mad
dogs and Englishmen.
(My thesis title: {\it Problems in the Classical and Quantum Theories of
Gravitation} was greeted
with hoots of derision when I announced it at the Cargese Summer School en
route to Trieste.
The work originated with a bet between Abdus Salam and Hermann Bondi about
whether you could
generate the Schwarzschild solution using Feynman diagrams. You can (and I did)
but I never found
out if Bondi ever paid up.)

 Inspired by Salam, Capper and I decided to use the recently discovered {\it
dimensional
regularization}\footnote {Dimensional regularization had just been invented by
another Salam student
and contempory of ours, Jonathan Ashmore \cite{Ashmore}, and independently by
Bollini and
Giambiagi \cite{Bollini} and by 't Hooft and Veltman \cite{thooft}.  I briefly
shared a London house
with Jonathan Ashmore and Fritjof Capra.  Both were later to leave physics, as
indeed was Derek
Capper. Ashmore went into biology, Capper into computer science and Capra into
eastern mysticism
(a decision in which, as far as I am aware, Abdus Salam played no part).} to
calculate corrections to
the graviton propagator from closed loops of massless particles: vectors
\cite{Capper1} and spinors
\cite{Capper2}, the former in collaboration with Leopold Halpern.  This
involved the self-energy
insertion
\begin{equation}
\Pi_{\mu\nu\rho\sigma}(p)=\int
{d^{n}xe^{ipx}<T_{\mu\nu}(x)T_{\rho\sigma}(0)>}|_{g_{\mu\nu}=\delta{\mu\nu}}
\label{selfenergy}
\end{equation}
where n is the spacetime dimension and $T_{\mu\nu}(x)$ the energy-momentum
tensor of
the massless particles. One of our goals was to verify that dimensional
regularization correctly
preserved the Ward identity
\begin{equation}
p^{\mu}\Pi_{\mu\nu\rho\sigma}(p)=0
\label{Ward}
\end{equation}
that follows as a consequence of general covariance.  If we denote by
$\epsilon$ the deviation from
the physical spacetime dimension one is interested in, and expand about
$\epsilon=0$, we obtain
\begin{equation}
\Pi_{\mu\nu\rho\sigma}=\frac{1}{\epsilon}\Pi_{\mu\nu\rho\sigma}(pole)+\Pi_{\mu\nu\rho\sigma}(finite)
\label{reg}
\end{equation}
Capper and I were able to verify that the pole term correctly obeyed the
Ward identity
\begin{equation}
p^{\mu}\Pi_{\mu\nu\rho\sigma}(pole)=0
\label{Ward2}
\end{equation}
and that the infinity could then be removed by a generally covariant
counterterm. We checked
that the finite term also obeyed the identity
\begin{equation}
p^{\mu}\Pi_{\mu\nu\rho\sigma}(finite)=0
\label{Ward3}
\end{equation}
and hence that there were no diffeomorphism anomalies.\footnote{Had we looked
at closed loops of Weyl
fermions or self-dual antisymmetric tensors in $2~ mod~ 4$ dimensions, and had
we been clever enough,
we would have noticed that this Ward identity breaks down.  But we didn't and
we weren't, so this
had to wait another ten years for the paper by Alvarez-Gaum\'{e} and Witten
\cite{Alvarez2}.}

 We were also aware that since the massless particle systems in question were
invariant under the
Weyl transformations \footnote{For fermions this is true $\forall n$; for
vectors only for $n=4$} of
the metric
\begin{equation}
g_{\mu\nu}(x)\rightarrow\Omega^2(x)g_{\mu\nu}(x)
\label{Weyl}
\end{equation}
together with appropriate rescalings of the matter fields, this implied that
the stress tensors in
(\ref{selfenergy}) should be traceless and hence that the self energy should
also obey the trace
identity
\begin{equation}
\Pi^{\mu}{}_{\mu\rho\sigma}(p)=0
\label{trace}
\end{equation}
We verified that the pole term was OK:
\begin{equation}
\Pi^{\mu}{}_{\mu\rho\sigma}(pole)=0
\label{trace2}
\end{equation}
consistent with our observation that the counterterms were not only generally
covariant but Weyl
invariant as well.  For some reason, however, I did not get around to checking
the finite term until
Christmas of 73 by which time I was back in England on my second postdoc, in
Oxford, where Dennis
Sciama was gathering together a group of quantum gravity enthusiasts. To my
surprise, I found that
\begin{equation}
\Pi^{\mu}{}_{\mu\rho\sigma}(finite)\neq0
\label{trace3}
\end{equation}
I contacted Derek and he confirmed that we hadn't goofed.  The Weyl invariance
(\ref{Weyl}) displayed
by classical massless field systems in interaction with gravity, first proposed
by Hermann Weyl in
1918 \cite{Weyl1,Weyl2,Weyl3}, no longer survives in the quantum theory! We
rushed off a paper
\cite{Capper3} to {\it Nuovo Cimento} (How times have changed!).

	I was also able to announce the result at The First Oxford Quantum Gravity
Conference, organised
by Isham, Penrose and Sciama, and held at the Rutherford Laboratory in February
74 \cite{Duff1}.
Unfortunately, the announcement was somewhat overshadowed because Stephen
Hawking chose the same
conference to reveal to an unsuspecting world his result \cite{Hawking} that
black holes evaporate!
Ironically, Christensen and Fulling \cite{Christensen1} were subsequently to
show that in two
spacetime dimensions the Hawking effect is  due entirely to the trace anomaly.
Two dimensional black
holes, and the effects of the Weyl anomaly in particular
\cite{Sanchez,Harvey,Jacobson}, are
currently enjoying a revival of interest.

\section{Anomaly? What anomaly?}
\label{Anomaly}
	Some cynic once said that in order for physicists to accept a new idea, they
must first pass
through the following three stages:

(1) It's wrong

(2) It's trivial

(3) I thought of it first.

In the case of the Weyl anomaly, however, our experience was that (1) and (2)
got interchanged.
Being in Oxford, one of the first people we tried to impress with our new
anomaly was J. C. Taylor
who merely remarked ``Isn't that rather obvious?''.  In a sense, of course, he
was absolutely
right.  He presumably had in mind the well-known result that theories which are
Weyl invariant in
n-dimensional curved space are automatically invariant under the conformal
group $SO(2,n)$ in the
flat space limit, which implies in particular that the dilatation current
$D^\mu\equiv x^\nu T^{\mu}{}_{\nu}$ is conserved:
\begin{equation}
\partial_{\mu} D^\mu = T^{\mu}{}_{\mu}+x^\nu\partial_{\mu} T^{\mu}{}_{\nu}
=T^{\mu}{}_{\mu} =0
\label{dilatation}
\end{equation}
Moreover, one already knew from the work of Coleman and Jackiw \cite{Coleman}
that such
flat-space symmetries suffered from anomalies. Of course, the Weyl invariance
(\ref{Weyl}), in
contrast to the conformal group, is a {\it local} symmetry for which there is
therefore no Noether
current. Nevertheless, perhaps one should not be too surprised to discover that
there is
a curved space generalization in the sense of a non-vanishing trace for the
stress tensor.
Consequently, Capper and I were totally unprepared for the actual response of
the rest of the
physics community: NO-ONE BELIEVED US!  To be fair, we may have put some people
off the scent by
making the correct, but largely irrelevant, remark that at one loop the
anomalies in the
{\it two-point} function could be removed for $n\neq2$ by adding finite local
counterterms
\cite{Capper3}. So we wrote another paper \cite{Capper4} stressing that the
anomalies were real and
could not be ignored, but to no avail. To rub salt in the wounds, among those
dismissing our result
as spurious were physicists for whom we youngsters had the greatest respect.
\newpage First the
Americans:

{\it...the finite $W_{reg}$ that is left behind by Schwinger's method, after
the infinities have been
split off, is both coordinate invariant and conformally invariant},  DeWitt
\cite{DeWitt}

{\it Something is wrong}, Christensen \cite{Christensen2}

{\it The form of the conformal anomaly in the trace of the stress tensor
proposed by a number of
authors violates axiom 5}, Wald \cite{Wald}

{\it Thus we find no evidence of the conformal trace anomalies reported by a
number of other
authors}, Adler, Lieberman and Ng \cite{Adler},

then the Europeans:

{\it Conformal anomalies in a conformally invariant theory do not arise...},
Englert, Gastmans and
Truffin \cite{Englert}

{\it There are important, physically relevant differences: most noticeable,
normalized
energy-momentum tensors do not possess a `trace anomaly'}, Brown and Ottewill
\cite{Brown},

especially the Russians:

{\it The main assumption of our work is that a regularization scheme exists
which preserves all the
formal symmetry properties (including the Weyl symmetry)...Therefore we hope
dimensional
regularization will give no anomalies..}, Kallosh \cite{Kallosh}

{\it It turns out that conformal anomalies, discovered in gravitating systems,
are not true
anomalies, since in modified regularizations they do not arise....
The above point of view on
conformal anomalies is shared by Englert et al.}, Fradkin and Vilkovisky
\cite{Fradkin1}

 {\it The presence or absence of the conformal anomaly depends on the choice
made between two
existing classes of covariant regularizations}, Vilkovisky \cite{Vilkovisky}

and, to be democratic, let us not forget the Greeks:

{\it The above method of regularization and renormalization preserves the Ward
identities...and
trace anomalies do not arise}, Antoniadis and Tsamis \cite {Antoniadis8}

{\it There exists a regularization scheme which preserves both general
coordinate invariance and
local conformal invariance (Englert et al)...}, Antoniadis, Iliopoulos and
Tomaras
\cite{Antoniadis1}

{\it In fact one needs a regularization scheme which preserves the general
coordinate and Weyl
invariance...such a scheme exists (Englert \footnote{If one starts with a
classically
non-Weyl invariant theory (e.g. pure Einstein gravity) and artificially makes
it Weyl invariant by
introducing via a change of variables $g'_{\mu\nu}(x)=e^{2\sigma
(x)}g_{\mu\nu}(x)$ an unphysical
scalar {\it spurion} $\sigma (x)$, then unitarity guarantees that no anomalies
can arise because this
artificial Weyl invariance of the quantum theory, $g'_{\mu\nu}(x) \rightarrow
\Omega ^2 (x)
 g'_{\mu\nu}(x)$ with $e^{2\sigma (x)}\rightarrow \Omega ^2 (x) e^{2\sigma
(x)}$, is needed to
undo the field redefinition and remove the spurious degree of freedom.
Professor Englert informed
me in Trieste that this is what the authors of \cite{Englert} had in mind when
they said that
anomalies do not arise.  Let us all agree therefore that many of the apparent
contradictions are due
to this misunderstanding.} et al)}, Antoniadis, Kounnas and Nanopoulos
\cite{Antoniadis2}.

 \section{London}
\label{London}
  As chance would have it, my third postdoc brought me to King's College,
London, at the same time as Steve Christensen, Paul Davies, Stanley Deser,
Chris Isham and Steve
Fulling. Bill Unruh was also a visitor.  It was destined therefore to become be
a hot-bed of
controversy and activity in Weyl anomalies. Provoked by Christensen and
Fulling, who had not yet
been converted, Deser, Isham and I decided to write down the most general form
of the trace of the
energy-momentum tensor in various dimensions \cite{Deser}.  By general
covariance and dimensional
analysis, it must take the following form: For n=2,
\begin{equation}
g^{\alpha\beta}<T_{\alpha\beta}> = aR,
\label{n=2}
\end{equation}
where $a$ is a constant.  For n=4,
\begin{equation}
g^{\alpha\beta}<T_{\alpha\beta}> = \alpha R^2+\beta R_{\mu\nu}R^{\mu\nu}+
\gamma
R_{\mu\nu\rho\sigma}R^{\mu\nu\rho\sigma}+\delta{}\raisebox{.7ex}{\fbox{}} R+c
F_{\mu\nu}{}^aF^{\mu\nu a},
\label{n=4}
\end{equation}
where $\alpha, \beta, \gamma, \delta$ and $c$ are constants. (In (\ref{n=4}) we
have allowed for
the possibility of an an external gauge field in addition to the gravitational
field.) For $n=6$,
$g^{\alpha\beta}<T_{\alpha\beta}>$ would have to be cubic in curvature and so
on. (At one-loop, and
ignoring boundary terms, there is no anomaly for n odd). I showed these
expressions to
Steve Christensen, with whom I was sharing an office, and asked him if he had
seen anything like
this before.  He immediately became very excited and told me that these were
precisely the
Schwinger-Dewitt $b_{n}$ coefficients. These are the $t$-independent terms that
appear in the
asymptotic expansion of the heat kernel of the appropriate differential
operators $\Delta$:
\begin{equation}
Tr~e^{-\Delta t}\sim \sum_{k=0}^{\infty} B_{k}
t^{(k-n)/2}~~~~~~~~~~~~~~t\rightarrow0^+
\label{kernel}
\end{equation}
where
\begin{equation}
 B_{k}=\int d^n x ~b_k
\label{B}
\end{equation}
For example, if $\Delta$ is the conformal Laplacian
\begin{equation}
\Delta= -\raisebox{.7ex}{\fbox{}}+\frac{R}{6}
\label{laplacian}
\end{equation}
then $b_4$ is given by
\begin{equation}
b_4=\frac{1}{180(4\pi)^2}[-R_{\mu\nu}R^{\mu\nu}+R_{\mu\nu\rho\sigma}R^{\mu\nu\rho\sigma}+
\raisebox{.7ex}{\fbox{}} R]
\label{b4}
\end{equation}
This was the road to Damascus for Steve as far as
Weyl anomalies were concerned and, like many a recent convert, he went on to
become their most ardent
advocate \footnote{ A delightful set of
reminiscences on the Weyl anomaly by Steve Christensen can also be found in
Bryce DeWitt's
festschrift \cite{Christensen3}.  During their stay at King's, he and Fulling
shared a flat in the
London borough of Ealing (home
of the famous {\it Ealing Comedy} movies). According to Christensen, the
connection
between trace anomalies and the Hawking effect occurred, Archimedes-like, to
Fulling while taking a
bath. He did not run through the streets of Ealing shouting ``Eureka'', but did
run upstairs to
the payphone to tell Bill Unruh.}. This was also the beginning of a very
fruitful collaboration
between the two of us.  The significance of my paper with Deser and Isham was
that, with the
exception of the ${\raisebox{.7ex}{\fbox{}}R}$ term in (\ref{n=4}), none of the
above anomalies
could be removed by the addition of finite local counterterms (hence the title
{\it Non-local
Conformal Anomalies}) and thus this laid to rest any lingering doubts about the
inevitability of
Weyl anomalies (or, at least, it should have done).  By this time, or shortly
afterwards, the
Hawking radiation experts at King's and elsewhere were arriving at the same
conclusion\cite{Christensen1,Davies1,Dowker1,Davies2,Bunch} as in fact was
Hawking himself
\cite{Hawking2}.  There followed a series of papers calculating the numerical
coefficients in
(\ref{n=2}) and (\ref{n=4}) and confirming that these were indeed just the
$b_{n}$ coefficients
\cite{Brown1,Brown2,Duncan,Dowker2,Tsao}.  These results were further
generalized to
self-interacting theories \cite{Drummond,Shore3,Shore4,Hathrell1,Hathrell2}.

One day about this time I answered the phone in my office only to hear those
five
words most designed to instill fear and trembling into the heart of a young
postdoc: ``Hi. This
is Steven Weinberg''.  Pondering on the non-renormalizability problem, Weinberg
had
become interested in quantum gravity in $2+\epsilon$ dimensions
\cite{Weinberg,Christensen4}.
Inspired by workers in statistical mechanics, who frequently work with
non-renormalizable field
theories but who nevertheless manage to extract sensible predictions, Weinberg
wondered whether this
might be true for gravity: was the theory ``asymptotically safe''? The answer
seemed to rely on the
{\it sign} of the two-dimensional trace anomaly i.e. on the constant $a$ in
(\ref{n=2}).
Accordingly, Weinberg set n=2 in the n-dimensional calculations of
\cite{Capper2} and concluded that
fermions had the wrong sign:
\begin{equation}
a = \frac{1}{24\pi}
\label{n=2trace}
\end{equation}
He repeated our calculation for scalars himself and found the same sign and
magnitude (consistent
with the observation that in two-dimensions there is a bose-fermi equivalence,
and consistent with
the black hole calculations \cite{Davies1,Christensen1}). In the case of vector
bosons, however, he
found from \cite{Capper1} that there was a sign flip.  His question was simple
but crucial: did I
agree with him or could there be an overall sign error? Not wanting to be the
victim of Weinberg's
wrath should I get it wrong, I spent several frantic days and sleepless nights
checking and
rechecking the calculations. Those who have ever chased a minus sign and those
who know Steve
Weinberg will appreciate my discomfort! In fact I agreed.  Unfortunately,
asymptotic safety became
asymptotically unpopular but my contact with Weinberg later led to a very
fruitful semester in
Austin, and to my continuing affection for the state of Texas.

The scalar terms of order $n/2$ in the curvature which appear in the
n-dimensional
gravitational trace anomaly are reminiscent of the pseudoscalar terms of order
$n/2$ in the
curvature which appear in the n-dimensional gravitational axial anomaly, as
calculated by Delbourgo
and Salam \cite{Delbourgo}.  I was musing on this shortly after moving across
town to Queen Mary
College, when I saw a paper by Eguchi and Freund \cite{Eguchi} on the then new
and exciting topic of
gravitational instantons.  They considered the two topological invariants, the
Pontryagin
number, $P$, and the Euler number, $\chi$, and posed the question: To what
anomalies do $P$ and
$\chi$ contribute?  In the case of the Pontryagin number, they were able to
answer this question by
relating $P$ to the integrated axial anomaly; in the case of the Euler number,
however, they found no
anomaly.  I therefore wrote a short note \cite{Duff5} relating $\chi$ to the
integrated trace
anomaly.  As described section \ref{string}, this result was later to prove
important in the
two-dimensional context of string theory, where
\begin{equation}
\chi=\frac{1}{4\pi}\int d^2x\sqrt{g}R
\label{chi}
\end{equation}
and hence from (\ref{n=2})
\begin{equation}
\frac{1}{ 4{\pi}}\int d^2x\sqrt{g}g^{\alpha\beta}<T_{\alpha\beta}> =  a{\chi}
\label{n=2euler}
\end{equation}
Unfortunately, the referee's vision did not extend
that far and the paper was rejected.  Rather than  resubmit it, I decided to
incorporate the results
into a larger paper \cite{Duff2} which re-examined the Weyl anomaly in the
light of its applications
to the Hawking effect, to gravitational instantons, to asymptotic freedom and
Weinberg's asymptotic
safety. In the process, I discovered that the constants $\alpha$, $\beta$,
$\gamma$ and $\delta$
are not all independent but obey the constraints
\begin{equation}
4\alpha+\beta=\alpha-\gamma=-\delta
\label{constraint}
\end{equation}
In other words, the gravitational contribution to the anomaly depends on only
two constants (call
them $b$ and $b'$) so that (\ref{n=4}) may be written
\begin{equation}
g^{\alpha\beta}<T_{\alpha\beta}> =b(F+\frac{2}{3}
\raisebox{.7ex}{\fbox{}}R)+b'G+cH
\label{bandb}
\end{equation}
where
\begin{equation}
F=R_{\mu\nu\rho\sigma}R^{\mu\nu\rho\sigma}-2R_{\mu\nu}R^{\mu\nu}+\frac{1}{3}R^2,
\label{F}
\end{equation}
\begin{equation}
G=R_{\mu\nu\rho\sigma}R^{\mu\nu\rho\sigma}-4R_{\mu\nu}R^{\mu\nu}+R^2
\label{G}
\end{equation}
and
\begin{equation}
H=F_{\mu \nu}{}^aF^{\mu \nu a}
\label{H}
\end{equation}
In four (but only four) dimensions
\begin{equation}
F=C^{\mu\nu\rho\sigma}C_{\mu\nu\rho\sigma}
\label{weyltensor}
\end{equation}
where $C_{\mu\nu\rho\sigma}$ is the Weyl tensor, and $G$ is proportional to the
Euler number density
\begin{equation}
G=^{*}R_{\mu\nu\rho\sigma}{}^{*}R^{\mu\nu\rho\sigma}
\label{eulerdensity}
\end{equation}
where $*$ denotes the dual. Note the absence of an $R^2$ term in (\ref{bandb}).
This result was later
rederived using the Wess-Zumino consistency conditions
\cite{Bonora,Pasti,Cappelli2,Osborn3,Cappelli1}.  Furthermore, the constants
$a$, $b$, $b'$ and $c$
are those which determine the counterterms  \begin{equation}
\Delta L=\frac{a}{\epsilon}\sqrt{g}
R~~~~~~~~~~~~~~~~~~~~~~~~~~~~~~~~~~~~~~~~~~~~~~~~~~~~~~~~~~~n=2
\label{n=2counterterm}
\end{equation}
\begin{equation}
\Delta
L=\frac{1}{\epsilon}\sqrt{g}(bF+b'G+cH)~~~~~~~~~~~~~~~~~~~~~~~~~~~~~~~~~~~~~~~~~~~~n=4
\label{n=4counterterm}
\end{equation}
(and hence the renormalization group $\beta$ functions) at the one-loop level.
The Euler number
counterterms are frequently ignored on the grounds that they are total
divergences, but will
nevertheless contribute in spacetimes of non-trivial topology. We emphasize
that the above results
are valid only for theories which are classically conformally invariant (e.g.
Maxwell/Yang-Mills
for n=4 only, and conformal scalars and massless fermions for both n=2 and
n=4).  For other
theories (e.g. Maxwell/Yang-Mills for n=2, pure quantum gravity for n=4, or any
theory with mass
terms) the ``anomalies'' will still survive, but will be accompanied by
contributions to
$g^{\alpha\beta}<T_{\alpha\beta}>$ expected anyway through the lack of
conformal invariance.  Since
the anomaly arises because the operations of regularizing and taking the trace
do not commute, the
anomaly in a theory which is not classically Weyl invariant may be defined as:
\begin{equation}
Anomaly~=g^{\alpha\beta}<T_{\alpha\beta}>_{reg}-<g^{\alpha\beta}T_{\alpha\beta}>_{reg}
\label{A}
\end{equation}
Of course, the second term happens to vanish when the classical invariance is
present.

 Note that in an expansion about flat space with
$g_{\mu\nu}=\delta_{\mu\nu}+h_{\mu\nu}$, $R$ is
$O(h)$, so it is sufficient to calculate the two-point function as in
\cite{Capper3} to fix the $a$
coefficient of $R$ in (\ref{n=2}). For n=4, $\raisebox{.7ex}{\fbox{}}R$ is
$O(h)$ while $F$
and $G$ are $O(h^2)$. Nevertheless, because of the
constraint (\ref{constraint}), a calculation of the two-point function is again
sufficient to
fix the $b$ coefficient of $C^{\mu\nu\rho\sigma}C_{\mu\nu\rho\sigma}$ in
(\ref{bandb}),
notwithstanding the ability to remove the $\raisebox{.7ex}{\fbox{}}R$ piece by
a finite local
counterterm, and notwithstanding some contrary claims in the literature.
Indeed, this is how the
coefficients of the Weyl invariant $C^{\mu\nu\rho\sigma}C_{\mu\nu\rho\sigma}$
counterterms were
first calculated \cite{Capper1,Capper3}.

Explicit calculations
\cite{Capper1,Capper3,Capper2,Deser3,Christensen1,Brown1,Bunch,Dowker2,Brown2,Duff2} yield
\begin{equation}
b=\frac{1}{120(4\pi)^2}[N_S+6N_F+12N_V]
\label{positive}
\end{equation}
\begin{equation}
b'=-\frac{1}{360(4\pi)^2}[N_S+11N_F+62N_V]
\end{equation}
where $N_S$, $N_F$ and $N_V$ are respectively the numbers of scalars, spin
$1/2$ Dirac fermions and
vectors in the theory. Note that the contributions to the $b$ coefficient are
all positive, as they
must be by spectral representation positivity arguments
\cite{Capper2,Deser3,Cappelli1} on the vacuum
polarization proper self-energy part in four dimensions.

 I also noted that the non-local effective action responsible for the
anomalies would contain a term \begin{equation}
S_{eff}\sim \int d^n x \sqrt{g} R ~\raisebox{.7ex}{\fbox{}}^{(n-4)/2} R
\label{Polyakov?}
\end{equation}
By setting $n=2$ one obtains what what later to be known as the {\it Polyakov
action}, discussed in
section \ref{string}.

\section{Cosmology}
\label{Cosmology}
The role of the Weyl anomaly in cosmology seems to fall into the following
categories: inflation
in the early universe, the vanishing of the cosmological constant in the
present
era, particle production and wormholes.

The first is reviewed in  papers by
Grischuk and Zeldovitch \cite{Grischuk} and Olive \cite{Olive}. Consider the
semi-classical Einstein
equations \begin{equation} R_{\mu\nu}-\frac{1}{2}g_{\mu\nu}R= 8{\pi}G
<T_{\mu\nu}>
\label{Einstein}
\end{equation}
where $<T_{\mu\nu}>$ is the effective stress-tensor induced by quantum loops.
In the inflationary phase, the geometry will be that of De Sitter space. But
the trace anomaly for De
Sitter completely determines the energy momentum because it must be a multiple
of the metric by the
symmetry \footnote{The trace anomaly also determines the energy momentum
completely for a two dimensional black hole and in the four dimensional case it
determines it up
to one function of position \cite{Christensen1}.}:
\begin{equation}
<T_{\mu\nu}>=\frac{1}{4}g_{\mu\nu}g^{\alpha\beta}<T_{\alpha\beta}>
\label{Desitter}
\end{equation}

	The idea that the trace anomaly might also have a bearing on the vanishing of
the cosmological
constant is a recurring theme
\cite{Christensen5,Antoniadis8,delAguila,Tomboulis,Antoniadis4,Antoniadis5,Antoniadis3}. According to
Tomboulis \cite{Tomboulis}, models where the cosmological constant relaxes
dynamically to zero via
some dilaton-like scalar field suffer from an unnatural fine-tuning of the
parameters.  This problem
can be cured, he claims, if the Wess-Zumino functional induced by the conformal
anomaly is included.
A similar approach has been taken by Antoniadis, Mazur and Mottola
\cite{Antoniadis4,Antoniadis5,Antoniadis3}, who argue that four-dimensional
gravity is drastically
modifed at distances larger than the horizon scale, due to the large infrared
quantum fluctuations
of the conformal part of the metric, whose dynamics is governed by an effective
action induced by
the trace anomaly, analogous to the {\it Polyakov action} in two dimensions.
Apparently, this leads
to a conformally invariant phase in which the effective cosmological term
necessarily vanishes.
See also \cite{Hawking3,Christensen5,Christensen6,Duff6} for a discussion of
the cosmological
constant/trace anomaly connection in the context of {\it spacetime foam}
\cite{Hawking3}.

With regard to particle production, Parker \cite{Parker} has used the trace
anomaly to argue that
there is no particle production by a gravitational field if spacetime is
conformally flat and
quantum fields are conformally coupled, but this has recently been challenged
by Massacand and
Schmid \cite{Massacand}.

Finally, Grinstein and Hill \cite{Grinstein} and also Ellis, Floratos and
Nanopoulos \cite{Ellis}
have claimed that in Coleman's wormhole scenario \cite{Coleman2}, it is the
trace anomaly that
controls the behavior of fundamental coupling constants, particle masses,
mixing angles, etc.

\section{Supersymmetry}
 \label{supersymmetry}

The Weyl anomaly acquires a new significance when placed in the context of
supersymmetry. In particular, Ferrara and Zumino \cite{Ferrara} showed that the
trace of
the stress tensor $T^{\mu}~_{\mu}$, the divergence of the axial current
$\partial_{\mu}J^{\mu 5}$,
and the gamma trace of the spinor current $\gamma^{\mu}S_{\mu}$ form a scalar
supermultiplet. There
followed a good deal of activity in calculating the corresponding anomalies in
global supersymmetry.

In the period 1977-9, Christensen and I found ourselves in Boston: he at
Harvard; I at Brandeis.  We
decided to look at these anomalies in supergravity. Since the supermultiplets
involve fields
$e_{\mu}~^a,\psi_{\mu},A_{\mu},\chi,\phi$ with spins $2,3/2,1,1/2,0$ we first
determined the axial
and trace anomalies for fields of arbitary spin
\cite{Christensen7,Christensen8}. See also
\cite{Hawking4,Yoneya,Townsend,Kallosh2,Nicolai}. One of our main motivations
was to calculate the
gravitational spin $3/2$ axial anomaly \footnote{Our result, that the
Rarita-Schwinger anomaly was
$-21$ times the Dirac, generated almost as much incredulity as did the
Capper-Duff Weyl anomaly, but
that's another festschrift.} which was at the time unknown.  Shortly afterwards
(by which time I had
come full circle and was back at Imperial College) van Nieuwenhuizen and I
noted \cite{Duff4} that
the gravitational trace anomaly for a field of given spin could depend on the
field representation.
Thus a rank two gauge $\phi_{\mu\nu}$ field yielded a different result from a
scalar $\phi$, even
though they are dual to one another.  Similarly, a rank three gauge field
$\phi_{\mu\nu\rho}$
yielded a non-zero result, even though it is dual to nothing.  However, these
differences showed up
only in the coefficient of the topological Euler number term.

At one loop in supergravity after going on-shell, we may write the anomaly as
\begin{equation}
g^{\alpha\beta}<T_{\alpha\beta}>=\frac{A}{32\pi^2}{}^{*}R_{\mu\nu\rho\sigma}{}^{*}R^{\mu\nu\rho\sigma}
\label{superA}
\end{equation}
so that when (\ref{bandb}) applies, $A=32\pi^2(b+b')$.
\begin{table}
$\begin{array}{crccc}
Field & 360A & N=8  & N=4  & N=4 \\
\bigskip
{}~ & ~ & supergravity & supergravity & Yang-Mills\\
e_{\mu}^a&848&1&1&0\\
\psi_{\mu}&-233&8&4&0\\
A_{\mu}&-52&28&6&1\\
\chi&7&56&4&4\\
\phi&4&63&1&6\\
\phi_{\mu\nu}&364&7&1&0\\
\bigskip
\phi_{\mu\nu\rho}&-720&1&0&0\\
{}~&~&A=0&A=0&A=0
\end{array}$
\caption{Vanishing anomaly in N=8 and N=4 supermultiplets.}
\end{table}
The contributions to the $A$ coefficient from the various fields are shown in
Table $1$. Note that the fermions are Majorana. The significance of these
results lies in their
application to the $D=4, N=8$ supergravity obtained by dimensional reduction
from Type II
supergravity in $D=10$ and the $D=4, N=4$ supergravity-Yang-Mills
supermultiplets which arise from
dimensional reduction from N=1 supergravity-Yang-Mills in D=10. As we can see
from field content
given in the last three columns of Table 1, the combined anomaly exactly
cancels \footnote{Curiously
enough, before the gravitino contribution to the anomaly was calculated
explicitly, D'Adda and Di
Vecchia \cite{Dadda} attempted to deduce it by assuming that the total anomaly
cancels in $N=4$
supergravity.  This was a good idea but they reached the wrong conclusion by
working with the dual
formulation with two $\phi$ fields instead of the stringy version with one
$\phi$ and one
$\phi_{\mu\nu}$ obtained by dimensional reduction.}.  I have singled out these
supermultiplets
because these are precisely the field theory limits of the toroidally
compactified Type II and
heterotic superstrings.  Indeed, these results have recently been confirmed in
a direct string
calculation by Antoniadis, Gava and Narain \cite{Antoniadis6}.

Another application of the trace anomaly in the context of supersymmetry
concerned the {\it gauged}
$N$-extended supergravities which exhibit a cosmological constant proportional
to the gauge
coupling $e$.  By calculating the Weyl anomaly in the presence of a
cosmological constant
\cite{Christensen5,Duff6}, therefore, one can determine the renormalization
group beta function
$\beta(e)$.  One finds, remarkably, that the one-loop $\beta$ function vanishes
for $N>4$
\cite{Christensen6}.

See \cite{Duff8} for a review of Weyl anomalies in supergravity, and
\cite{Fradkin2} for those in
conformal supergravity.
\section{The string era}
\label{string}

The history of the Weyl anomaly took a new turn with the advent of string
theory. The emphasis
shifted away from four dimensional spacetime to the two dimensions of the
string worldsheet.  In
particular, in two very influential 1981 papers, Polyakov
\cite{Polyakov1,Polyakov2} showed that
the critical dimensions of the string correspond to the absence of the two
dimensional Weyl anomaly.
In the first quantized theory of the bosonic string, one starts with a
Euclidean functional integral
over the worldsheet metric $\gamma_{ij}[\xi]$, $i,j=1,2$, and spacetime
coordinates $X^{\mu}[\xi]$,
$\mu=0,1,...,D-1$, where $\xi^i$ are the worldsheet coordinates. Thus
\begin{equation}
e^{-\Gamma}=\int \frac{D\gamma~DX}{Vol(Diff)} ~e^{-S[\gamma,X]}
\end{equation}
where
\begin{equation}
S[\gamma,X]=\frac{1}{4\pi\alpha'}\int d^2 \xi
\sqrt{\gamma}\gamma^{ij}\partial_iX^{\mu}\partial_jX^{\nu}\eta_{\mu\nu}
\end{equation}
As showed by Polyakov, the Weyl anomaly in the worldsheet stress tensor is
given by
\begin{equation}
\gamma^{ij}<T_{ij}>=\frac{1}{24\pi}(D-26)R(\gamma)
\end{equation}
The contribution of the D scalars follows from (\ref{n=2trace}) while the $-26$
arises from the
diffeomorphism ghosts that must be introduced into the functional integral.  In
the case of the
fermionic string, the result is
\begin{equation}
\gamma^{ij}<T_{ij}>=\frac{1}{16\pi}(D-10)R(\gamma)
\end{equation}
Thus the critical dimensions $D=26$ and $D=10$ correspond to the preservation
of the two
dimensional Weyl invariance $\gamma_{ij}\rightarrow\Omega^2(\xi)\gamma_{ij}$.
One may wonder how
Polyakov addressed the controversy of the previous eight years described in
section 3. Well he
didn't, but merely remarked ``This is the well-known trace anomaly''.

Previously, the critical dimensions had been understood from the central charge
$c$ of the
Virasoro algebra \cite{Virasoro}
\begin{equation}
[L_n,L_m]=(n-m)L_{n+m}+\frac{c}{12}n(n^2-1)\delta_{n,-m}
\end{equation}
where the $L_n$ are the coefficents in a Laurent expansion of the stress
tensor, namely
\begin{equation}
T(z)=\sum_{n \in Z} L_n z^{-n-2}
\label{laurent}
\end{equation}
where $T=T_{zz}$ and $z\equiv exp(\xi^0+i\xi^1)$. Thus this established a
connection between the
two dimensional Weyl anomaly and the central charge of the Virasoro algebra (to
be precise,
 $c=24\pi a$ in
equation (\ref{n=2})) ; a result which
spawned the whole industry of {\it conformal field theory} in the context of
strings. See, for
example, Alvarez-Gaum\'{e} \cite{Alvarez}.  In fact, when writing
(\ref{laurent}), one usually
assumes that $T_{ij}$ is traceless, which forces the anomaly to show up as a
diffeomorphism anomaly,
but the results are entirely equivalent \cite{Green}.

Polyakov went on to describe what happens in {\it non-critical} string theory
when the Weyl
invariance is lost, and the metric conformal mode propagates.  In this case,
the two dimensional
effective action is given by
 \begin{equation}
S_{eff}\sim\int d^2 \xi \sqrt{\gamma} [R ~\raisebox{.7ex}{\fbox{}}^{-1} R+\mu]
\end{equation}
where we have allowed for a worldsheet cosmological term produced by quantum
corrections. If we
now separate out the conformal mode $\sigma$ and let $\gamma_{ij}\rightarrow
e^{\sigma}\gamma_{ij}$, we obtain the Liouville action
\begin{equation}
S_L[\sigma]=\int d^2 \xi \sqrt{\gamma}
(\frac{1}{2}\gamma^{ij}\partial_i\sigma\partial_j\sigma
+R\sigma+\mu e^{\sigma})
\end{equation}
This is the starting point for much of non-critical string theory.

The role of the Weyl anomaly becomes even more interesting when we allow for
the presence of the
spacetime background fields, as shown by Callan et al \cite{Callan} and Fradkin
and Tseytlin
\cite{Fradkin3}.  In the case of the bosonic string, for example, the
worldsheet action takes the
form

\[
S=\frac{1}{2\pi\alpha '}\int d^2 \xi
\frac{1}{2}[\sqrt{\gamma}\gamma^{ij}\partial_iX^{\mu}\partial_jX^{\nu}G_{\mu\nu}(X)
+\epsilon^{ij}\partial_iX^{\mu}\partial_jX^{\nu}B_{\mu\nu}(X)] \]
\begin{equation}
+\frac{1}{4\pi}\int d^2 \xi\sqrt{\gamma}R(\gamma)\Phi(X).
\label{S}
\end{equation}
corresponding to background fields $G_{\mu\nu}(X)$, $B_{\mu\nu}(X)$ and
$\Phi(X)$.  Now the
anomaly may be written as
\begin{equation}
\frac{1}{2\pi\alpha'}\gamma^{ij}<T_{ij}>=\beta^{\Phi}\sqrt{\gamma}R(\gamma)+\beta^{G}{}_{\mu\nu}\sqrt{\gamma}\gamma^{ij}
\partial_iX^{\mu}\partial_jX^{\nu} +\beta^{B}{}_{\mu\nu}\epsilon^{ij}
\partial_iX^{\mu}\partial_jX^{\nu}
\end{equation}
The absence of the Weyl anomaly thus means the vanishing of the $\beta$
functions, which to lowest
order turn out to be \cite{Callan}
\begin{equation}
0=\beta^{G}_{\mu\nu}=R_{\mu\nu}-\frac{1}{4}H_{\mu}{}^{\lambda\sigma}H_{\nu\lambda\sigma} + 2
\nabla_{\mu}\nabla_{\nu}\Phi + O(\alpha ')
\end{equation}
\begin{equation}
0=\beta^{B}_{\mu\nu}=\nabla_{\lambda}H_{\mu\nu}{}^{\lambda}-2\nabla_{\lambda}\Phi
H^{\lambda}{}_{\mu\nu} + O(\alpha ')
\end{equation}
\begin{equation}
0=16\pi^2\beta^{\Phi}=\frac{D-26}{3\alpha'}+[4(\nabla \Phi)^2 - 4 \nabla^2\Phi-
R-\frac{1}{12}H^2] + O(\alpha '^2)
\end{equation}
But these are nothing but the Einstein-matter field equations that result from
the action
\begin{equation}
\Gamma_{eff}\sim\int d^D x\sqrt{-G}
e^{-2\Phi}[\frac{D-26}{3\alpha'}-R-4(\nabla\Phi)^2-\frac{1}{12}H^2]+...
\end{equation}
The common factor $e^{-2\Phi}$ reveals that these terms are tree-level in a
string loop perturbation
expansion. If we denote the dilaton vacuum expectation value by $\Phi_0$, then
from (\ref{chi}) the
classical action (\ref{S}) yields a term
\begin{equation}
e^{-\chi\Phi_0} =e^{-2(1-L)\Phi_0}
\end{equation}
in the functional integral, where $L$ counts the number of holes in the
two-dimensional Riemann
surface, i.e. the number of loops.

That the Einstein equations in spacetime should originate from the vanishing of
the worldsheet
Weyl anomaly is perhaps the most remarkable result in our story.

\section{Current Problems}
\label{current}

New techniques for calculating Weyl anomalies continue to appear. Fujikawa
\cite{Fujikawa1,Fujikawa2} has pioneered the functional integral approach where
the origin of the
lack of quantum Weyl invariance in a classically invariant theory may be
attributed to a
non-invariant measure in the functional integral.  Ceresole, Pizzochero, and
van Nieuwenhuizen
\cite{Ceresole} have reproduced these results
using flat-space plane waves.  Bastianelli and van Nieuwenhuizen
\cite{Bastianelli1,Bastianelli2}
have applied to Weyl anomalies the quantum mechanical approach, first used by
Alvarez
Gaum\'{e} and Witten \cite {Alvarez2} in the context of axial anomalies.

Much of the current interest in the Weyl anomaly resides not only in
high-energy physics and general
relativity but in statistical mechanics.  See, for example
\cite{Alvarez,OAlvarez,Friedan1,Friedan2,Affleck,David,Polchinski}. A
particularly powerful result
is Zamolodchikov's $c$-theorem \cite{Zamolodchikov}, which states that there
exists a function
defined on the space of two-dimensional conformal field theories which is
decreasing along
renormalization group (RG) trajectories, and is stationary only at RG fixed
points, where its value
equals the Virasoro central charge $c$. Recently, there has been a good deal of
activity by Cardy
\cite {Cardy}, Osborn \cite{Osborn}, Jack and Osborn \cite{Jack}, Cappelli,
Friedan and Latorre
\cite{Cappelli1}, Shore \cite{Shore1,Shore2}, Antoniadis, Mazor and Mottola
\cite{Antoniadis7} and
Osborn and Petkos \cite{Osborn2} attempting to generalize this theorem to
higher dimensions. Whereas
from (\ref{n=2}) the two dimensional gravitational anomaly depends on only one
number, however, from
(\ref{n=4}) the four dimensional one depends on two: the Euler term and the
Weyl term. In higher
dimensions, there will always be one Euler term, but the number of Weyl
invariants grows with
dimension \cite{Deser2}.  Consequently, it is not clear whether there is a
unique way to generalize
the theorem nor how useful such a generalization might be.  Cardy and Jack and
Osborn focussed on the
Euler number term, whereas Cappelli et al pointed to the positivity of the Weyl
tensor term
(\ref{positive}) as a more likely guide, but the analysis is still
inconclusive.

On the subject of the Euler term, the results of section \ref{supersymmetry}
are, in fact, still
controversial because the anomaly inequivalence between different field
representations, discussed by
van Nieuwenhuizen and myself, was challenged at the time by Siegel
\cite{Siegel} and by Grisaru,
Siegel and Zanon \cite{Grisaru}.  They found that the traces of the two stress
tensors were
equivalent . Yet the recent string results of Antoniadis et al
\cite{Antoniadis7} would seem to
support our interpretation. Moreover, if it were incorrect, the vanishing of
the anomalies for the
N=4 and N=8 multiplets would seem to be a gigantic coincidence.

Nevertheless, to tell the truth (in accordance with Salam's maxim), I am still
uneasy about the whole
thing. The numerical coefficients quoted in Table 1 were calculated using the
$b_4$ coefficients
discussed in section $3$.  The claim that these correctly describe the trace
anomaly is in turn
based on an identity which everyone used to take for granted.  See, for
example, Hawking
\cite{Hawking2}. The identity says that if $S[g]$ is a functional of the metric
$g_{\mu\nu}$, then
\begin{equation}
\int d^{n}x ~g^{\mu\nu}\frac{\delta S[g]}{\delta g_{\mu\nu}} \equiv
\frac{\partial
S[\lambda g]}{\partial \lambda}|_{\lambda=1}
\label{identity}
\end{equation}
where $\lambda$ is a constant. If true, it would mean that the integrated trace
of the stress
tensor arises exclusively from the non-invariance of the action under constant
rescalings of the
metric.  However, the {\it Polyakov} action provides an obvious counterexample:
\begin{equation}
S_{eff}\sim\int d^2 x \sqrt{g} R ~\raisebox{.7ex}{\fbox{}}^{-1} R
\label{Polyakov}
\end{equation}
This action is scale invariant, but gives rise to an anomaly proportional to
$R$! Of course, the
integrated anomaly is a purely topological Euler number term, and it is the
topological nature of
the action which provides the exception to the rule. However, since the entire
debate over
equivalence versus inequivalence devolved precisely on Euler number terms, the
use of the
identity (\ref{identity}) in this context makes me feel very uneasy.

The whole question of whether the dual formulations of supergravity yield the
same Weyl anomalies has
recently been thrust into the limelight with the string/fivebrane duality
conjecture
\cite{Duff7,Strominger}
which states that in their critical spacetime dimension $D=10$, superstrings
(extended objects with
one spatial dimension) are dual to superfivebranes (extended objects with five
spatial dimensions
\cite{Bergshoeff}).  Whereas the two dimensional worldsheet of the string
couples to the rank two
field $\phi_{\mu\nu}$, the six dimensional worldvolume of the fivebrane couples
to the rank six field
$\phi_{\mu\nu\rho\lambda\sigma\tau}$.  The usual rank two formulation of D=10
supergravity
dimensionally reduces to the $N=4$ field content of Table 1, but the dual rank
six formulation
reduces to a different field content with $20~
\phi_{\mu\nu\rho}$ and with $14 ~\phi$ replaced by $14~ \phi_{\mu\nu}$. If
Table 1 is to be
believed, $A(\phi_{\mu\nu})-A(\phi)=1$ and $A(\phi_{\mu\nu\rho}) =-2$.
Therefore the dual version has
non-vanishing coefficent $A=14-40=-26$. This increases my uneasiness.

A possible resolution of this problem may perhaps be found in the recent paper
by Deser and
Schwimmer\cite{Deser2} who have re-examined the different origin of the
topological versus Weyl
tensor contributions to the anomaly (which they call Type A and Type B,
respectively).

I certainly believe that the final word on this subject has still not been
written.

\section{Acknowledgements}
\label{ack}
I would like to thank my anomalous collaborators Derek Capper, Steve
Christensen, Stanley Deser,
Chris Isham, and Peter van Nieuwenhuizen.

Most of all, however, my thanks go to my former supervisor Abdus Salam who
first kindled my
interest in quantum gravity and who continues to provide inspiration to us all
as a scientist and
as a human being. When the deeds of great men are recalled, one often hears the
clich\'{e} ``He
did not suffer fools gladly'', but my memories of Salam at Imperial College
were quite the
reverse. People from all over the world would arrive and knock on his door to
expound their latest
theories, some of them quite bizarre.  Yet Salam would treat them all with the
same courtesy and
respect.  Perhaps it was because his own ideas always bordered on the
outlandish that he was so
tolerant of eccentricity in others; he could recognize pearls of wisdom where
the rest of us saw
only irritating grains of sand. As but one example of a crazy Salam idea, I
distinctly remember him
remarking on the apparent similarity between the mass and angular momentum
relation of a Regge
trajectory and that of an extreme black hole.  Nowadays, of course, string
theorists will juxtapose
black holes and Regge slopes without batting an eyelid, but to suggest this
back in the late 1960's
was considered preposterous by minds lesser than Salam's.\footnote{Historical
footnote: at the time
Salam had to change the gravitational constant to match the hadronic scale, an
idea which spawned his
{\it strong gravity}; today the fashion is the reverse and we change the Regge
slope to match the
Planck scale!}

Theoretical physicists are, by and large, an honest bunch: occasions when
scientific
facts are actually deliberately falsified are almost unheard of. Nevertheless,
we are still
human and  consequently want to present our results in the best possible light
when writing them up
for publication. I recall a young student approaching Abdus Salam for advice on
this ethical
dilemma: ``Professor Salam, these calculations confirm most of the arguments I
have been making so
far. Unfortunately, there are also these other calculations which do not quite
seem to fit the
picture. Should I also draw the reader's attention to these at the risk of
spoiling the effect or
should I wait? After all, they will probably turn out to be irrelevant.'' In a
response which should
be immortalized in {\it The Oxford Dictionary of Quotations}, Salam replied:
``When all else fails,
you can always tell the truth''.

Amen.

\pagebreak

\end{document}